\documentclass[letterpaper,twocolumn,10pt]{article}
\usepackage{usenix,epsfig,endnotes}
\usepackage{supertabular}
\usepackage{epstopdf}
\usepackage{graphicx}
\usepackage{url}
\usepackage{seqsplit}
\begin{document}

\date{}

\title{\Large \bf An Empirical Study of Mobile Ad Targeting}

\author{
{\rm Theodore Book}\\
Rice University
\and
{\rm Dan S. Wallach}\\
Rice University
} 

\maketitle


\subsection*{Abstract}
Advertising, long the financial mainstay of the web ecosystem, has become nearly ubiquitous in the world of mobile apps.  While ad targeting on the web is fairly well understood, mobile ad targeting is much less studied.  In this paper, we use empirical methods to collect a database of over 225,000 ads on 32 simulated devices hosting one of three distinct user profiles.  We then analyze how the ads  are targeted by correlating ads to potential targeting profiles using Bayes' rule and Pearson's chi squared test.  This enables us to measure the prevalence of different forms of targeting.  We find that nearly all ads show the effects of application- and time-based targeting, while we are able to identify location-based targeting in 43\% of the ads and user-based targeting in 39\%.

\section{Introduction}

1.3 million of the 1.5 million apps available on Google's Play Store are available for free, including 99\% of the apps with at least 50,000 downloads~\cite{appBrainFreeVsPaid}.  This is largely made possible by advertising, which accounts for the majority of mobile app revenue in the United States, and reached USD 19.3B world-wide in 2013~\cite{ventureBeatMonetizationTypes, iabAdSpend2013}.  Other forms of app revenue, such as paid app sales and in-app purchases, make up the rest of the mobile app monetization space.  This huge market depends on delivering ads to people who are likely to respond to them.

Mobile ads are delivered through advertising libraries which are embedded in the application at compile time.  These libraries generally use an embedded web browser, Android's WebView, to display the ads and handle user interaction.  (These WebViews are subject to many of the same security concerns that impact other dated web browsers~\cite{danWebViewTinker}.)  These libraries then handle the work of requesting the ad from a central server, displaying it to the user, and tracking the user's interaction with the ad.

In prior work, researchers have measured the ability of Android advertising libraries to access permission-protected user data~\cite{grace2012unsafe, stevensinvestigating, leontiadis2012don,book2013longitudinal} as well as the behavior of applications that directly pass user private information to their ad libraries~\cite{book2013collusion}. What remains little understood is the way that information is used after it has been collected.

In this work, we measure a more complex factor that is also critical to the understanding of user privacy---the interaction between advertising libraries and their host servers.  Because we do not have direct access to the proprietary algorithms used in processing ad requests and serving advertisements, we choose to treat the data center as a black box, observing the relationship between inputs (ad requests) and outputs (provided ads).  This resembles the methodology used by other researchers in measuring the targeting of web applications~\cite{lecuyer2014xray, barford2014adscape, carrascosa2014understanding, olejnik2014selling}.

In-app mobile ads present interesting opportunities for ad targeting, and bring related concerns regarding user privacy.  Advertisers face the task of delivering their ads to a receptive audience.  Some forms of Internet advertising, such as search advertising, allow effective targeting without any user profile at all---search keywords, alone, can indicate products and services for which a user might be shopping.  Other forms of advertising, such as on social networks, rely on extensive profiles that the user has entrusted to the service, giving the service provider the responsibility of using the data for ad targeting without revealing its contents to third parties~\cite{dave2014computational, korolova2010privacy}.

In-app mobile ads, by contrast, present relatively little information that advertisers can use for targeting.  At their most basic level, they provide a prospective advertiser with little more than the name of a host application and an IP address that can be used for rough geographical targeting.  Because of this limitation, advertising libraries have long used some form of unique identifier to identify the device, and hence, the user~\cite{book2013longitudinal}.  When unique identifiers are correlated with other data---passed by the application or obtained from other sources---opportunities for more precise ad targeting begin to emerge.  Our goal is to measure and quantify the targeting that takes place, with an eye to elucidating the privacy implications that stem from that targeting.

In order to maintain a reasonable scope for our work, we chose to focus on a single platform (Android), and a single ad library (Google's AdMob.)  Android represents 84.4\% of current mobile device shipments, making it a good proxy for the entire mobiles space~\cite{idcq4marketshare}.  Likewise, AdMob represents 35\% of the install-weighted ad library market on the Android platform, making it a reasonable proxy for the entire mobile ad space~\cite{book2013collusion}.  It is worth noting that AdMob is one of the more conservative advertising libraries in regard to the amount of user data that it collects~\cite{book2013longitudinal}.  Other libraries, with access to more personal data, may engage in additional forms of targeting.

In studying AdMob, we are not only studying the behavior of the Google's proprietary systems.  Rather, we are studying the entire AdMob ecosystem, which includes not only Google, with its engineers and systems, but also the advertisers that make use of those systems to place ads based on various criteria.  This includes many third parties that use ad exchanges, such as Google's DoubleClick, to place ads on the AdMob network.  In this way, our research represents an attempt to yield an end-to-end understand of a complex multi-vendor system.

\section{Methodology}

Measuring mobile ads presents methodological challenges due to the need to measure an ever changing data set without the ability to take random samples~\cite{guha2010challenges}.  New ad campaigns are constantly being introduced, old ones retired, and ads targeting affects each request.  We sought to overcome these factors by focusing on narrow segments of mobile ads, with a limited number of profiles, over a restricted time period.  In order to generate requests, collect ads, and correlate responses to the requests, we first reverse engineered the AdMob ad request protocol.  Following this, we built an AdMob emulator that would request ads based on app parameters that we supplied.  We then developed a collection of user profiles that would enable us to measure ad targeting.  We next rolled out a collection infrastructure consisting of numerous collection stations that reported ads back to our central server.  Finally we analyzed the data we collected to determine the correlation between possible targeting methodologies and the ads that we received.

\subsection{Reverse Engineering AdMob}
In order to give ourselves maximal flexibility in probing AdMob's behavior, we chose to reverse engineer the AdMob ad request API rather than send requests from stock applications on physical devices.  This allowed us the freedom of manipulating various parameters and observing the results.  We began by capturing traces of the AdMob library's communications on a number of real and emulated devices.  Because these exchanges are unencrypted, it was easy for us to analyze them and identify critical components.~\footnote{It also makes it easy for attackers to make use of identifying information in the request to profile individuals.  Indeed, it is known that nation-state attackers have made use of this information in the past~\cite{snowdenMobileLocation}.}

By using a variety of applications in generating our requests, we were able to capture the behavior of various versions of the AdMob Android library, and thus build a fairly complete understanding of its behavior.  While the AdMob library makes a variety of HTTP requests in the course of its operation, we found that all of the data necessary to request an ad is passed in a single HTTP GET request~\cite{vallina2012breaking}.  

We analyzed our request dataset and identified 88 different parameters set by the AdMob library.  Our next task was to determine the significance of each of those parameters in order to emulate the library's behavior.  We proceeded by decompiling the AdMob library and examining the code used to generate the request.  In this way, we were able to identify the code used to set the parameter, and identify the sources of the information that was encoded.

While most of the parameters related to the mechanics of requesting and serving ads, some parameters with significance for user privacy are listed in Table~\ref{tab:params}.  As AdMob uses ordinary HTTP to make requests and serve ads, the content of most of these parameters is sent in clear text.  The one exception is the UULE parameter, which transmits the device location, and is encrypted with a key embedded in the AdMob library.

\begin{table}
  \centering
\begin{tabular}{l p{5cm}}
\textbf{Field} & \textbf{Explanation} \\
\hline
an &  App name and version \\
cap &  List of device capabilities.\\
                 &         m - maps\\
                &          a - apps (handles market URLs)\\
                 &         t - make telephone calls\\
carrier &  The numeric name (MCC+MNC) of the network operator. \\
client &  Adsense publisher account ID\\
cust\_age &  App supplied user age\\
cust\_gender &  App supplied user gender\\
gnt &  Network type from Android telephony manager \\
hl &  System language\\
isu & First 32 characters of an MD5 hash of the Android ID \\
kw  &  App supplied keywords \\
slotname & Identifies unique ad space \\
u\_audio &  Current state of audio playback\\
& Speakerphone on / audio playing\\
& Ringer on / vibrate / silent\\
u\_tz &  Timezone in minutes from GMT\\
uule  &  Encrypted location\\

\end{tabular}
\caption{Some privacy related AdMob request parameters}
\label{tab:params}
\end{table}

\subsection{Building an AdMob Emulator}
After reverse engineering the AdMob request syntax it became possible to develop an AdMob emulator, which would build requests that would be interpreted as legitimate by AdMob servers.  We constructed the emulator in Python, taking application characteristics as inputs, and generating syntactically valid AdMob requests.  The completed emulator allowed us to vary parameters related to ad targeting while continuing to generate correct values for the built-in checksum and other changing parameters.  We tested our emulator to verify that it retrieved valid ads from the AdMob network.

Of course, the URL parameters are not the only pieces of information that AdMob servers are able to gather from a request.  The request timing and IP address are also significant.  Because of this, it was necessary to develop an infrastructure that would enable us to send requests from various IP addresses with a timing corresponding to a realistic refresh rate for a mobile ad library.  We resolved this challenge by building a client server infrastructure where clients running on multiple machines would poll a central server to request a set of ad request URLs.  Each client provides a unique ID to which the server assigns a specific profile.  The client then requests ads from AdMob at a regular interval, returning the results to the server for later analysis.

\subsection{Device Simulation}

AdMob requests include a unique device ID---either a hash of the Android ID, used on older versions, or the Android Advertising ID, used on newer versions.  We used this phenomenon to create virtual ``devices,'' each with a unique ID.  We assigned each simulated device to a different client, with only one device per client.  In this way, we were able to isolate the devices from each other.

In order to ensure our simulated devices did not receive similar ads because they were all located on the same university network, we ensured that many of them were located on private residential Internet connections in the central area of our city.  We avoided using generally available proxies such as Tor or PlanetLab nodes, or resources such as Amazon S3, under the assumption that those IP addresses might trigger fraud detection mechanisms at Google, or at the very least would be grouped differently from IP addresses used by actual user devices.  Indeed, preliminary testing on public proxies generally resulted in an empty response from Google---a phenomenon that was associated with rejected requests.

The selection of physical locations was also important.  By keeping them all in the same geographic area, we were able to minimize ad differences based on large scale ad targeting.  This helped us to ensure that the difference between the ads received on different devices were due to their different user profiles, and not to different geographic locations.  Nevertheless, we were able to explore some aspects of geographic targeting within our small collection area.

In this way, we constructed a network of 32 simulated devices, requesting ads for a variety of applications from the AdMob network.  Each device was configured to represent a unique user ``profile'' defined mainly by the apps that were installed on the simulated device.  Each had a ``fresh'' randomly generated device ID that should not have had any prior data associated with it.

While we attempted to control for factors other than our user profiles that could be used in targeting ads, it should be noted that our selection of IP addresses may have had some effect on targeting.  Not only may these IP addresses have had slightly different geographical mappings, they may also have been associated by AdMob with previous or simultaneous traffic by the owners of the address.  This may have resulted in ads targeted due to factors other than our user profile.  We sought to control for this by assigning a number of devices to each user profile, allowing us to control for variations consistent with prior data associated with a given IP address. (Our home users were generally connected to the Internet with DSL or cable modems, using dynamically assigned IP addresses. External, long-term observers would presumably not be able to do much targeting unless they had more information from the ISP about the customer identity behind the IP address. 

\subsection{User Simulation}

Profiles were constructed by querying actual users from different demographic segments about the applications that they used on their mobile devices.  This information was supplemented with ``recommended applications'' lists for various demographic groups obtained from the Internet.  Once lists of applications were prepared, we harvested the advertising identifiers from the applications in question.  Because AdMob is now included in the Google Play Services library incorporated in many Android applications, it was not sufficient to simply test for the presence of AdMob in the compiled application.  Rather, we ran the applications on physical Android devices, and captured network traces for all connections with the Internet.  We then analyzed the traces to extract any requests to AdMob, and used our knowledge of the AdMob request syntax to extract the identifiers necessary to simulate requests from the app.  In this way, we verified that the applications we selected were actively displaying AdMob ads.

In the process, we also captured traffic from a number of other advertising libraries, which were often much less well behaved than AdMob.  We observed libraries transmitting lists of installed applications, user city, state and zip code, and other personal information, all in clear text.  However, as our research focused on AdMob targeting, we did not seek to further quantify these observations.

\begin{table}
\small
\centering
\begin{tabular}{p{2.3cm} p{2.3cm} p{2.3cm}}
\textbf{Profile 1} & \textbf{Profile 2 }& \textbf{Profile 3} \\
\hline
Fly Dragon, Fly! & Realtime Stock Quotes & Police Car Driver 3D \\
\small{\textit{com.\allowbreak lsgvgames.\allowbreak slideandfly}} & \small{\textit{org.\allowbreak dayup.\allowbreak stocks}} & \small{\textit{com.\allowbreak ScnStudios.\allowbreak Traffic\allowbreak Police\allowbreak Car\allowbreak Driving}}  \\
\hline
Happy Fall & ConvertPad--Unit Converter & Drag Racing \\
\small{\textit{com.\allowbreak noodlecake.\allowbreak happyfall}} & \small{\textit{com.\allowbreak mathpad.\allowbreak mobile.\allowbreak android.\allowbreak wt.\allowbreak unit}} & \small{\textit{com.\allowbreak creativemobile.\allowbreak DragRacing }}\\
\hline
Happy Jump & King James Bible & Trains and Friends \\
\small{\textit{com.\allowbreak noodlecake.\allowbreak happyjump}} & \small{\textit{com.\allowbreak hmobile.\allowbreak biblekjv}} & \small{\textit{com.\allowbreak puzzletopstoday.\allowbreak trainpuzzleforkids}} \\
\hline
Tiny Flashlight & Tiny Flashlight & Tiny Flashlight \\
\small{\textit{com.\allowbreak devuni.\allowbreak flashlight}} & \small{\textit{com.\allowbreak devuni.\allowbreak flashlight}} & \small{\textit{com.\allowbreak devuni.\allowbreak flashlight}} \\

\end{tabular}
\caption{Profiles Used in Testing}
\label{tab:profiles}
\end{table}

We developed a pool of candidate applications from the applications where we were able to capture AdMob advertising information.  We only included apps that make exclusive use of the leaderboard format (where ads are displayed as a small banner together with other content).  This helped us to ensure that the ads shown across different applications were comparable.

We selected apps from this pool to form user profiles---each of which was a list of apps that would be requested on a device assigned that profile.  While it would have been possible to construct a nearly unlimited number of such profiles, we settled on three, a number that allowed sufficient contrast to measure the degree to which targeting occurred, while maximizing the sample size for each profile.  In this way, we were able to focus our limited data collection capacity on a small number of profiles, enhancing our ability to differentiate between ads targeted at one profile or another.

Likewise, in order to keep our data as focused as possible, we chose only three apps and a control for each profile, ensuring that we captured the maximum number of ads for each app within the constraints of our limited collection network, thus minimizing accidental variance between apps.  Table~\ref{tab:profiles} shows the apps associated with each profile.  For a control application, we selected a flashlight application with a very high number of installs (over 100 million).  We made this decision  based on the assumption that a broadly used application like a flashlight would not strongly influence any user profile in which it was included.  This was intended to ensure that the ads served to our different profiles would primarily be influenced by the applications that differed among them, and not by the control.  At the same time, having the flashlight application in all of our profiles enabled us to measure whether the same ads were served to the flashlight app regardless of the profile, or whether the other apps in the profile influenced the ads served to the flashlight app.

We designed our simulated requests for each app to be as close as possible to the actual requests generated by the app.  For most apps, this consisted of providing strings such as an app id and client id, but two apps also included a static list of key words with each request.  The \textit{Realtime Stock Quotes} app included the following static string: \texttt{finance, funds, loans, mortgage, stock}, while the \textit{Fly Dragon, Fly!}, app included the following string: \texttt{Dragon Fly, Tiny Wings, Android, game, Dragon, entertainment, casual}.  We included these strings in our simulated requests.

\subsection{Network and Data Capture}

Having constructed our model and our collection apparatus, we proceeded to collect data for a period of one month, coinciding with December of 2014.  December is an important month for retail advertisers, making it a logical candidate.  While a shorter period would have better enabled us to avoid noise due to ad campaigns commencing and terminating, our limited collection apparatus required a longer period of time to gather a rich enough collection of ads for thorough analysis.  While we attempted to maintain all of our collection points in action throughout the entire month, we experienced some downtime at approximately half of our collection sites, largely due to the vagaries of using personal machines on private networks.  Nonetheless, we managed to collect 225,000 individual ads directing users to 5,735 different URLs.

For each ad, we recorded the time the ad was received, the simulated device it was received on (which mapped to a specific user profile), the app it was received for, and the raw contents of the ad.  Each ad consisted of a HTML document, which generally referenced outside images and scripts.  We developed infrastructure to parse the various ad formats that we received and extract salient features.  As is common in web advertising, many of the ads that we received used JavaScript redirects to display content from another source~\cite{springborn2013impression}.  In this case, we downloaded the redirect target and stored it in the database, as well.  Frequently, embedded scripts were used with the same function---the actual content of the ad was loaded by an external script.  For this reason, we downloaded and stored all scripts.  Rather than downloading static images, we simply stored the image URL.

Our primary factor for differentiating ads was the ad's target, by which we mean the URL to which the user would be redirected if they clicked on the ad.  We extracted the ad target from the final redirect.  When recording the target URL, we removed cache busting parameters and the like, so that all URLs that targeted a given page were counted together.  In many cases, the target was in the form of a URL that linked to one site, while redirecting to another site embedded in a URL parameter.  In these cases, we took the final site as the target.  In less than 1\% of our ads, we were not able to discern the final site due to a redirect format that varied for each ad, and which did not embed a redirect URL.  In these cases, we used the image URL as a proxy for the final destination.

A curious question is whether our actions here corresponded to users clicking on ads, which might cause downstream ``pay-per-click'' cashflows, and which could well have brought our measurement infrastructure to the attention of AdMob's internal clickfraud infrastructure. We believe, by extracting the final URL from its redirector without actually visiting the redirector itself, that we avoided generating anything that the AdMob infrastructure would recognize as a click. 

\subsection{Data Analysis}

We used two primary statistical methods in analyzing our data according to our specific needs.  When we desired to see if ads were uniformly distributed across various groups, we made use of Pearson's chi-squared test~\cite{pearson1900x}.  This gives us the probability that the distribution of ads between two groups is due to chance, and thus, the probability that factors other than a uniform distribution resulted in the direction of certain ads to certain groups.\footnote{A group might be all ads corresponding to a user profile, app, or device, for example.}  In other words, if a feature influenced the targeting of ads, we would expect to see a high probability (generally greater than 99\%) that the distribution of ads across the various groups was not random.  If a feature did not not produce targeting, we would expect to see a lower probability, as the distribution of ads approached a uniform distribution.

When it came to quantifying the correlation between a specific feature (such as a user profile) and the presence or absence of specific ads, we made use of Bayes' rule.  This enabled us to quantify the probability that a given group was present based on the presence of the ad.  In other words, it gave us the correlation of a given ad with a given group.  Combined with the known frequency of each group in the data set, this enabled us to estimate the degree to which an ad was targeted towards a specific profile.  In other words, if an ad was targeted at a particular group with no other relevant factors, we would expect to see a 100\% correlation between the ad and the group.  If there was no targeting involved, we would expect to see the correlation be equal to the group's share of the total population.

Both of these methods require a sufficient sample size to work properly.  When we have a single ad for a given advertiser, it is impossible to infer anything useful about its distribution among our categories.  For this reason, we chose to only analyze ads for which we had at least 50 impressions.  This enabled us to have a confidence interval of 8.3\% or less on results in the 90th percentile with 95\% surety.  That is to say that if we estimated that 90\% of the impressions for a certain ad appeared on a specific profile, we would be 95\% certain that our estimate was correct within plus or minus 8.3 percentage points.  While we could have selected other values, we believe that this cut-off allows us to confidently identify where targeting occurs, despite allowing minor uncertainty in the exact level of targeting that may have been affected a particular ad.

\section{Findings}

We were able to reach a number of interesting conclusions through our analysis of our data.  It appears that most ads in our dataset were targeted through automated means.  Nearly all ads showed evidence of being targeted at the application level.  43\% of our ads were geographical in nature, but it appeared that targeting was not accurate beyond the level of the metropolitan area.  95\% of ads showed time-based targeting effects, as well, although it is possible that the intentional targeting of some ads skewed the time distribution of others.   Finally, we were able to determine that 39\% of the ads for our control application showed the influence of user-based targeting.  More interestingly, we were able to identify a number of ads that appeared to be targeted at the human users of the IP addresses on which we were conducting our testing.

\subsection{Immediate Observations}
\begin{figure}
\centering
\includegraphics[width=0.5\textwidth]{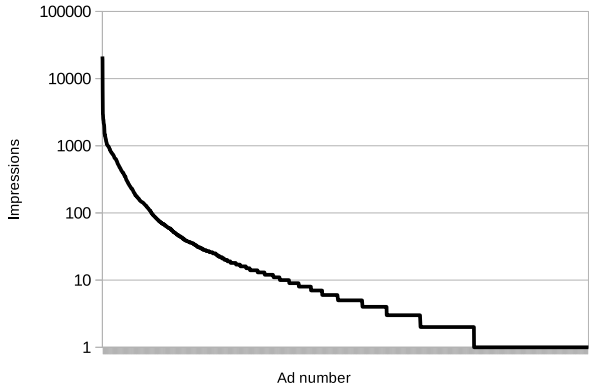}
\caption{Number of impressions per ad}
\label{fig:impressions}
\end{figure}


\begin{table}
  \centering
\begin{tabular}{r r}
\textbf{Number of Networks Detected} & \textbf{Number of Ads} \\
\hline
0	&19734 \\
1	&128112 \\
2	&79637 \\
3	&1075 \\
4	&236 \\

\end{tabular}
\caption{Number of ad networks detected found for different ads}
\label{tab:redirects}
\end{table}

We found 3,078 unique targets out of our dataset of 228,794 ads.  Some targets were extremely common, with 21,372 impressions recorded for the real estate website Trulia.com.  At the other end of the spectrum, we received only a single impression for 725 targets.  See figure~\ref{fig:impressions}.  The large number of unique targets shows the vibrancy of the mobile advertising market.

We also tallied the total number of distinct advertising networks that were involved  for our ads.  We extracted network names from the JavaScript that loads the ads, as well as the click URLs.  Often, a single ad might involve the coordination of several ad networks.  For example, a click URL might include a parameter that provides the redirect URL for another ad network, and that URL might, in turn, point to a third.  Of course, our analysis was limited by our ability to identify network involvement from the HTML, JavaScript, and target URLs that we received.  We may have failed to identify some networks whose activity did not leave any trace in our record, our when our analysis software was simply not calibrated to detect the traces of involvement from a given network.

We were able to identify 27 distinct ad networks that were involved in placing ads on AdMob.  Some ads had multiple networks.  The number of networks that we were able to identify for each ad are listed in table~\ref{tab:redirects}.  The fact that multiple networks are involved sometimes points to the mergers in the advertising industry (for example, the acquisition of JumpTap by Millenial Media)~\cite{jumptapAcquired}.  However, it also points to the nature of ad exchanges, where companies can purchase ads on each other's networks.  We will discuss some of the privacy implications of these exchanges in section~\ref{exchanges}.

Before entering into a statistical analysis of the data, it is worth noting some patterns that emerge from the ad URLs, themselves.  Often, we see some identifying data encoded into the ad request URL, allowing the advertiser to track or target ads based on some features of the device.  For example, car ads from everesttech.net feature strings like \texttt{ev\_pl=\allowbreak mobileapp:\allowbreak :2-com.\allowbreak puzzletopstoday.\allowbreak trainpuzzleforkids} and \texttt{ev\_dvm=android+sdk}.  These strings, of course, tell us little about the targeting in the delivery of the ads, but only about customization and tracking that might occur after a user clicks on an ad.

Other clues provide some glimpses into the targeting mechanisms used.  For example, on-click URLs in ads from Luminosity feature strings such as \texttt{Category=arcade}, \texttt{Category=tools}, or \texttt{Category=finance}, which correspond to the app in which the ad is embedded (slideandfly, mathpad, and DayUp Stocks, respectively).  This suggests that that the ads have been targeted at specific categories of apps.  (They all redirect to the same URL.)

In other cases, multiple ads by the same advertiser can give us some hints into targeting.  For example, four distinct Twitter ads were found with significant frequency in our dataset.  One redirected to the main Twitter homepage, one to US news, one to US sports, and one to a Tim Howard page.\footnote{Tim Howard, goalie for the U.S. side in the recent World Cup competition.}  All four ads appeared only on the Trains and Friends children's game app, but they were evenly spread across the different devices requesting ads for that app.  This app, which involves assembling a 9-16 piece puzzle of a popular toy train, appears to be targeted at children around the age of 3, which is not the usual Twitter demographic. In this case, ads that seem to be designed for different demographics appear to have all received the same incorrect targeting---all being displayed on an app targeted at toddlers too young to use Twitter.  This implies a failure in an automated targeting system.  Presumably Twitter created the different ads to appeal to different demographics, and relied on an automated targeting system---either its own, Google's, or one belonging to a third party ad broker.  This system determined that users of the Trains and Friends app were likely to respond to Twitter ads---perhaps because of a click or two by a toddler who liked pretty blue birds.  This might have directed additional Twitter ads to the same, presumably inappropriate, app.  Of course, other scenarios could explain this coincidence, but the use of automated techniques to select the app in question seem likely.

Sometimes, however, different campaigns from the same advertiser target different apps.  For example, three Dollar General campaigns in our dataset appeared on different apps.  \textit{The Easy Meals Holiday Blogger Challenge} ads displayed broadly across the children's apps and flashlight app.  The \textit{Gifts that Glow} campaign appeared exclusively on the \textit{Tiny Flashlight} app across all of our devices, while the \textit{Tips and Ideas} campaign appeared exclusively on the Happy Fall and Happy Jump applications on two of the devices that requested ads for those applications.

The broad spread of requests across different devices using these applications suggests that we are dealing with application-based targeting.  However, the question remains of whether a human with a sense of humor decided to target \textit{Gifts that Glow} at a flashlight application, or whether an algorithm discerned that flashlight users might be attracted to glowing things.  In order better to understand these questions, let's consider some empirical results.

\subsection{App-Based Targeting}

\begin{figure}
\centering
\includegraphics[width=0.5\textwidth]{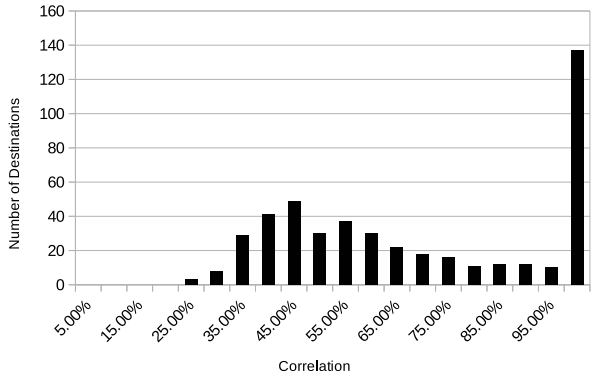}
\caption{Histogram showing correlation of ads to apps}
\label{fig:app-histogram}
\end{figure}

It is no surprise that ads are targeted on the basis of which app that they are displayed in.  What is more interesting is that every ad in our dataset showed a statistically significant correlation with the app in which it was displayed.  All ads showed a greater than 0.9 cumulative probability, with the great majority showing a greater than 0.999 CP of app-based targeting based on Pearson's chi squared test.  Indeed, 125 of the 466 ads with at least 50 impressions appeared in only one application.  When the distribution of ads across apps is plotted in a histogram, an interesting bimodal distribution appears.  See figure~\ref{fig:app-histogram}, where each ad is counted based on the app to which it has the highest correlation.  138 of the ads show a greater than 95\% correlation with a specific app.  The rest are spread in a wide distribution centered on 40-45\% correlation.  This is a higher correlation than one would expect in a random distribution across 10 apps, which explains why the chi squared test showed that all of the apps showed the effects of app-based targeting.

The 45\% mode implies that the ads that are not exclusively shown on a single app are still more likely to appear on some apps than others.  That is to say that there is a group of one or more apps where the given ad is particularly likely to appear.  While some of these groupings may be due to ads being redistributed due to the app-based targeting of other ads, we will show in the following sections that a great deal of the effect is due to user and device-based targeting.

There are three possible sources of app-based targeting.  It could originate with the advertisers themselves selecting certain apps to target.  It could emerge from intermediaries who place ads on behalf of the advertisers, or it could emerge from the design of the AdMob system, itself.  It is likely that some of the distribution of ads across apps are due to the limitations of our collection model such as the extended collection period.  Nonetheless, given that no ads within our dataset were uniformly distributed across all of our applications it appears that some methodology was applied to all of them.  As AdMob was the only entity involved in the placement of every ad in our dataset, it seems highly probable that AdMob uses application-based targeting in directing ads.

Of course, this does not exclude the likelihood of targeting by advertisers and other intermediaries, nor does it exclude the use of manual targeting.  Indeed, the AdWords campaign management console (which is used to place AdMob ads) includes app specific settings~\cite{adwords-console}.  However, the incidence of humorously misdirected ads, such as Twitter ads being sent to toddlers and weekend car rental ads being sent to drag racing game players, suggests that many ads were targeted to a particular application based on an automated system rather than human involvement.


\subsection{Location-Based Targeting}

\begin{figure}
\centering
\includegraphics[width=0.5\textwidth]{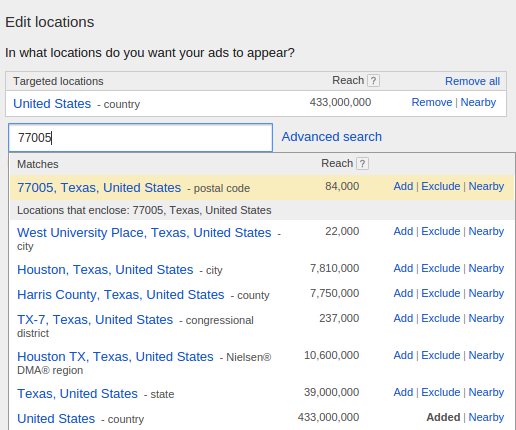}
\caption{AdWords console---Location targeting}
\label{fig:adwords-location}
\end{figure}

Location-based targeting is also possible with varying degrees of precision.  It can be as general as targeting a metropolitan region, or as specific as targeting ads only to users standing outside a specific store.  The AdWords campaign management console allows location-based targeting down to the neighborhood level~\cite{adwords-console}.  See figure~\ref{fig:adwords-location}.  Our methodology sought to minimize opportunities for location-based targeting by locating all of our collecting stations in a limited geographic area.  Nonetheless, we were able to extrapolate some data regarding location-based targeting from our data set.

While it was impossible for us to extract advertisers' actual targeting settings from AdMob, we were able to gather a great deal of information from the content of their ads.  We manually surveyed all of the click URLs that appeared at least 50 times in our dataset, identifying those intended for a local audience.  We based this determination on whether the ad was for a local business, or made an explicit reference to our location.  In this way, we determined that at least 43\% of our ads were geographically targeted.  Of course, this number represents a lower bounds, as ads without any explicit regional references may still have been part of a regional advertising campaign.

We then sought to understand how accurate the targeting was.  A few ads were wildly mis-targeted.  For example, we received 520 impressions for the Salem, Mass. Planet Fitness location, and 55 impressions for a Renton, Washington Ford dealership---both over 24 hours away from our location by car.  Aside from a few such aberrations, however, most of the targeted ads correctly identified at least our metropolitan area.

One factor in our dataset enabled even more detailed analysis.  Included in our dataset were 137 ads, totaling 65,827 impressions, from localsaver.com.  These ads had the the convenient URL format of localsaver.com/[STATE]/[CITY]/[BUSINESS]/[AD].  This enabled us to know which locations were being targeted by the advertiser, at least at the city level.  A localsaver.com customer support representative clarified to us that ads would only be shown within a ten mile radius of a business location.

Out of the localsaver.com ads that we received, 99 were targeted at the core city where our collection stations were located.  The remaining 38 were targeted at seven suburban locations that surrounded the central city, one smaller city 90 miles away, and one location that appeared to be mis-classified from a nearby state.  Only one of the suburban locations was within a ten mile radius of any of our researchers.

Assuming that there was no intentional misdirections of ads targeted for one area to a nearby neighborhood, this implies that the location-based data that Google was able to extract from our requests was accurate only to the level of our metropolitan area.  We chose not to provide GPS coordinates with our requests, as the majority of apps with the AdMob library enabled do not collect GPS data~\cite{book2013longitudinal, book2013collusion}.

To sum up, 43\% of our ads were geographical in nature.  We were able to identify that a large portion of these (at a minimum, the localsaver.com ads) were targeted at more specific locations within our metropolitan area.  However, there was no evidence that Google was able to reliably identify our location with a greater precision than at the metropolitan level.

\subsection{Time-Based Targeting}
\begin{figure}
\centering
\includegraphics[width=0.5\textwidth]{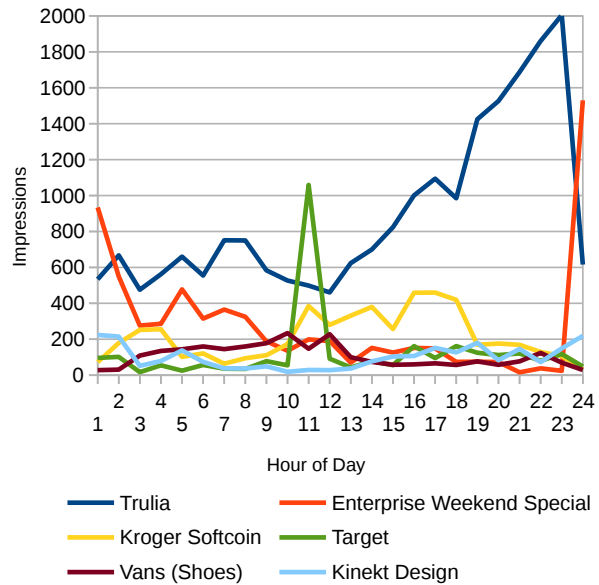}
\caption{Time distribution of six most common ads}
\label{fig:hourly}
\end{figure}


Another dimension of ad targeting practiced by advertisers and enabled by Google is time-based advertising~\cite{adwords-console}.  While the user's timezone is the only personal data needed to practice this sort of advertisement, it allows advertisers to target a user at specific points of their day, and thus, implicitly, at different locations.  An ad displayed in the middle of the day may well be seen at work, one displayed in the evening will likely be displayed at home.

In order better to understand this phenomenon, we created 24 buckets, one for each hour of the day, and examined the distribution of ads for each target within those 24 buckets.  This enabled us to measure the distribution of our ads across the different times of the day.  Figure~\ref{fig:hourly} shows the time distribution for the 6 most heavily advertised sites in our dataset.

As can be seen, there are notable time-based factors in some of these ads.  The Trulia ad is shown much more heavily in the afternoon and evening, with a sharp drop-off at midnight.  By contrast, the Enterprise ad spikes at midnight, and trails off through the early hours of the morning.  The Target ad has a curious spike at 11:00 AM, while the Kroger Softcoin ad is most heavily displayed in the afternoon.

While some of the variations may have been due to oddities in the distribution system --- for example, spending a daily budget that began at midnight --- others seem to imply intentional targeting.  The Trulia curve, for example, would suggest an advertiser choosing to target individuals browsing the Internet after work---perhaps a likely time for individuals to explore real estate.

We know that individual advertisers can prioritize certain times of day for their advertisements.  Less well known is whether the AdMob system uses automated time-based targeting algorithms.  Nearly all of our advertisements showed strong time-based variations, with 95\% of ads showing a less than 1\% probability of having a uniform distribution across the various hours of the day, according to Pearson's chi squared test.  It is possible that this is an artifact of certain ads competing for various prominent hours of the day, and other ads being forced to less desirable times.

To better understand this phenomenon, we considered the ``early morning'' ads---those which had the majority of their impressions shown between midnight and 6 AM.  We identified 58 ads with 14,000 impressions as being in that category.  We then categorized these ads in an attempt to see if there was anything in common that would suggest an algorithm targeting late night users.  However, the ads were broadly spread across a variety of categories, without any obvious correlation to late night activity.  This suggested that their prevalence in the late night category was simply due to pressure from other advertisers in other time slots rather than any human or algorithmic targeting.  However, this conclusion must be regarded as tentative, as it is difficult to account for all of the factors that might be used for targeting.

\subsection{User Based Targeting}

\begin{figure}
\centering
\includegraphics[width=0.5\textwidth]{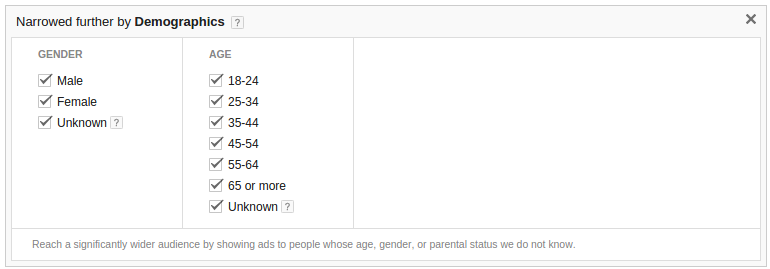}
\caption{AdWords console---Demographic targeting}
\label{fig:adwords-demographics}
\end{figure}

\begin{figure}
\centering
\includegraphics[width=0.5\textwidth]{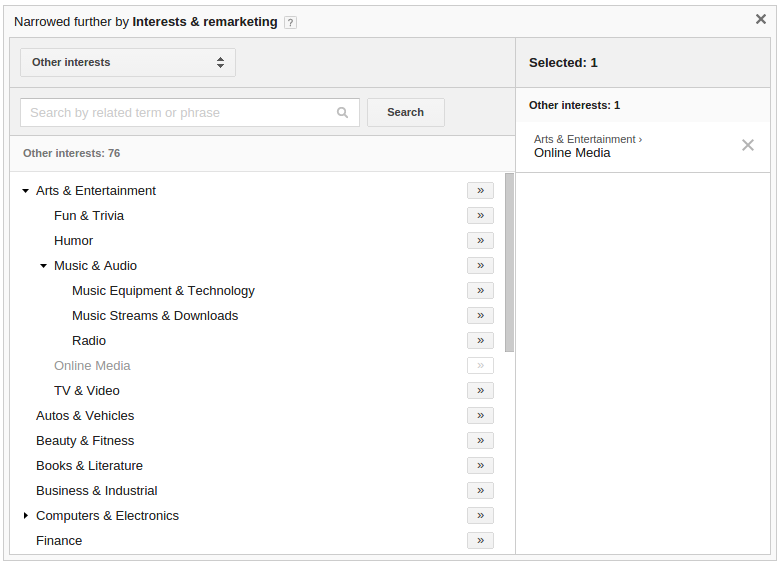}
\caption{AdWords console---Interest targeting}
\label{fig:adwords-interests}
\end{figure}

Perhaps the most powerful and potentially intrusive form of targeting is targeting based on the actual user of the device.  To the extent that the advertiser is able to make use of a user profile that is predictive of interests, motivations, and purchasing habits, they will be able to direct ads to users who are most likely to be influenced by them.

Our system developed user profiles built around collections of applications---a single ``device'' had a few applications installed on it, which a simulated user used in sequence.  One application---a flashlight application---was the same across all simulated devices.  Our hypothesis was that the AdMob infrastructure would build a profile of user interests based on the installed applications, and display different ads in the flashlight application based on the different user profiles.  To the extent that the ads are random, targeted at the flashlight app, or targeted by factors such as location or time (which are similar across all devices), we would expect a uniform distribution of ads across profiles.  To the extent that the ads are targeted based on user profiles, we would expect the profile to correlate with certain advertisements.

We applied Bayes' theorem to all of the ads that were displayed at least 50 times for the flashlight app, calculating the probability that a given user was requesting ads given that a a specific ad was being displayed.  We found that 77 of the 199 ads (39\%) predicted a given user with greater than 90\% certainty.  The fact that a sizable number of ads correlated with a given profile is sufficient to indicate that an ad may be targeted based on prior applications used on the same device.

\begin{figure}
\centering
\includegraphics[width=0.5\textwidth]{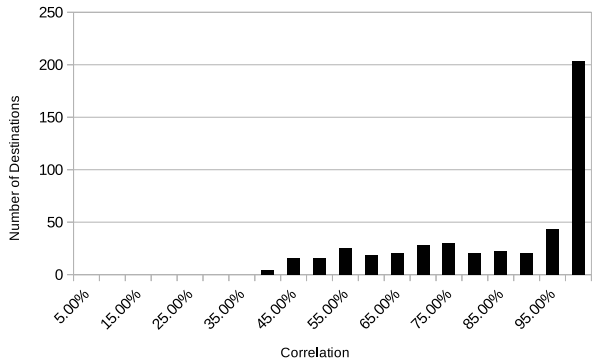}
\caption{Histogram showing correlation of ads to profiles}
\label{fig:profile-histogram}
\end{figure}

Looking at the broader group of all ads, we see that 116 ads appear exclusively on a single profile.  This is actually lower than the 125 ads that target a single app, due to the flashlight app being included in multiple profiles.  However, we find that 203 ads show at least a 95\% correlation with a single profile, as opposed to the 137 that showed the same level of correlation with a single app.  Figure~\ref{fig:profile-histogram} is a histogram showing the correlation of different ads to specific profiles.  It lacks the clear bimodal character of figure~\ref{fig:app-histogram}, while generally showing a high level of correlation.

It should be noted that this figure subsumes many of the effects of app-based targeting, as most apps are included in only a single profile.  At the same time, it helps us to understand how ads that appear in multiple apps are targeted based on user profiles.  The shift from a bimodal to unimodal distribution implies that much of the grouping of ads across a small subset of applications was in fact due to user targeting.  The smaller number of ads shown exclusively on a single profile combined with the larger number of ads that were at least 95\% predictive of a single profile shows that there may be somewhat less accuracy to user targeting, as a few ads are targeted outside the profile, while the great majority reach the same profile.  Finally, it should be noted that any device-based targeting will also appear as profile-based targeting in this data.

\subsection{Targeting based on real users}

\begin{table}
  \centering
\begin{small}
\begin{tabular}{p{2.3cm} p{2.3cm} p{2.3cm}}
\textbf{Device 1} & \textbf{Device 2}& \textbf{Device 3} \\
\hline
Maid Service & Electronic Test Equipment & Martial Arts \\ 
Blogger Challenge & New Homes & Cosmetic Dentistry\\
The Bath Specialist & Brain Games & Shredding\\
Home Staging & Casino & Insurance\\
Vans (shoes) &  MS in Technology Commercialization & Business Financing\\
Japanese Art Auction & ChromeCast & Business Law\\
Christian High School & BlackJack & Enterprise Backup\\
Spa Services &  & Student Loans\\
Church &  & Advertise Your Business\\
High-end appliances &  &

\end{tabular}
\end{small}
\caption{Ads that appear to have been IP targeted}
\label{tab:user-targeting}
\end{table}

Surprisingly, even after accounting for user profile and application, there were some significant variations across devices.  Some of this may be explainable by fine grained geographic targeting that differentiated among our device locations, and some may be explainable by the limitations of our data collection methods.  Nonetheless, there is reason to believe that some of this variation may be due to other Internet traffic from the same home network or IP address.

To be clear, a substantial portion of the advertisements observed on a given profile showed an approximately uniform distribution across the various devices that employed the profile.  (cumulative probability \textless~0.75)  Nonetheless, there were certain advertisements that appeared nearly exclusively on specific devices.  Table~\ref{tab:user-targeting} shows ads that appeared on selected devices more than twice as frequently as they appeared on other devices with the same profile.

After sharing these lists with the individuals that agreed to the use of their home networks for data collection, several of the individuals were able to confirm that the lists appeared to be related to their Internet browsing.  This seems to indicate that the variance is due to user targeting, and not to geographic variations, natural variance, or timing issues.

It is also noteworthy that certain devices attracted a large number of ads not seen on any other device, while others remained much closer to the median for their profile.  This further suggests that for certain devices (particularly the ones on home networks) additional targeting information was available, resulting in advertisements that differed from the norm.

The method used in targeting these ads is not entirely clear.  It seems probable that they are targeted based on other traffic from the same network or device. Our collector ran as a Python script on a host computer that may have also been used for web browsing, but did not make use of any browser components, eliminating the possibility of shared cookies or other features that might identify a particular browser instance.

There are multiple methods that could be used to identify a network or device in a semi-persistent manner.  One of the most invasive would be something similar to Verizon's PrecisionID---a system that attaches tracking metadata to all requests from a specific mobile device~\cite{bergenPrecisionId}.  However, the authors of this paper are unaware of any such system for the home DSL and cable networks with dynamic IP addresses, such as the ones used for the majority of our collection stations.  This suggests IP address-based targeting.

Because most home Internet systems do not provide for static IP addresses, IP-based targeting by itself would not provide a way to persistently associate behavior with a subgroup of users.  Two techniques could be used by AdMob to overcome this deficiency.  The simplest would be to maintain IP-based targeting data for a limited period of time, associating only recent behavior with a given IP address.  A more invasive system would associate an IP address with a user profile when identifying information is sent from that address---for example, a cookie in a web browser.  Once the association is made, ads could be targeted to the IP address based on information previously collected on that user, without a need for identifying information to be included in every request.  After a timeout, or if the user appeared at a different address, the association could be removed.

\paragraph*{Privacy Implications}

Assuming that our testing apparatus received ads that were targeted based on other traffic from the same network---whatever methodology was used to characterize the users of that network---several privacy concerns emerge.  One concern is that advertisements can be used to infer the targeting criteria used to display the advertisements in question~\cite{lecuyer2014xray, castelluccia2012betrayed}.  This means that anyone with access to a network can infer characteristics of the other individuals using that network.  If the number of users is relatively small, it may well be possible to attribute those characteristics to a specific individual.  Another concern relates to the databases used to target the ads.  To the extent that individual users are identified, a profile of their usage habits and behavior can be assembled that could imply various pieces of personal or medical data that are associated with significant privacy concerns.

\subsection{Targeting and Ad Exchanges}
\label{exchanges}

So far, we have distinguished between different entities that may engage in ad targeting.  Now, we would like to focus on the specific mechanisms that allow organizations other than Google to target ads on AdMob.  AdMob participates in Google's DoubleClick ad exchange~\cite{adMobOnDoubleClick}.  Advertising exchanges allow advertisers, brokers, and other advertising networks to bid on ad inventory in real time, and provide various data points that bidders can use in targeting their ads.  The DoubleClick ad exchange interface allows a bidder to decrypt the Android Advertising ID (a unique device ID).  It also provides some or all of the following data: the first three bytes of the user's IP address, a numerically coded location at the city and zip code level, an application identifier, a geofence of latitude and longitude coordinates, user time zone, gender, age range, user membership in cookie-defined remarketing lists, user languages, content classification as weighted values on up to 2208 codes, mobile device platform, brand, model and OS version,  mobile carrier, screen size and orientation, user rating for the app displaying the ad, as well as information regarding ad format~\cite{doubleClickProtocol}.

As can be seen, this information is more than sufficient to target ads without relying on any targeting algorithms employed by Google.  It is also difficult to detect which ads were directly placed with AdMob, and which have been brokered through DoubleClick.  Given this limitation, the infrastructure about which we are drawing our conclusions must be thought of in a comprehensive sense, as including not only Google's own infrastructure, but the infrastructure of other companies that place ads through DoubleClick.

\paragraph*{Privacy Implications}

The practice of sharing ad request data with third parties through ad exchanges exposes a number of privacy related questions.  While the request APIs are not open, and a significant amount of server capacity would be needed to process the full data stream, it exposes sensitive user information to a potentially unlimited number of third parties.  The combination of geographic information and a unique device ID makes it possible to associate the ID with a specific individual.  (Device-unique IDs are also available through API calls by Android apps with only the default permissions.)  It then becomes possible to track that individual throughout their day by observing their change in location and which apps they use, with all of the data that can be gleaned by knowing at what times of the day a user accesses their phone.

Because the consumers of the ad exchange data are themselves advertisers, it would be reasonable to expect that they would make use of this data---either now or in the future---to integrate the available data with outside data sets in order to build more comprehensive user profiles and hence better marketing.  The possibilities become even greater when one realizes that ad exchanges receive data from multiple ad libraries on multiple platforms, and thus have the potential to provide a very comprehensive view of a vast number of people.

\section{Comparison with Web Advertising}

Given that web advertising is relatively well understood, and that the social consensus regarding privacy issues in web advertising is better formed, it is worthwhile to compare targeting in mobile applications to targeting in web advertisements.  First of all, it is worth noting that the two forms of advertising have much in common---mobile ads consist of HTML rendered in a compact web browser embedded in an application.  Furthermore, we have observed that the distribution mechanisms for mobile ads resemble those for web advertisements.

Targeted ads based on user profiles are ubiquitous in the web environment, despite various policy initiatives to limit them~\cite{mayer2012third}.  Indeed, successful companies have built their business models around building user profiles for the purpose of ad targeting~\cite{blueKai}.  Likewise, researchers have made progress in measuring ad targeting and determining the criteria used to target different ads~\cite{lecuyer2014xray}.

As with web advertising, mobile ads are not necessarily hosted by the ad network that displays them.  Rather, redirects and mediation responses direct the advertising library to display ads from a variety of sources over the standard HTTP interface.  This allows additional parties exposure to the AdMob library, and allows them to practice their own targeting.

Despite the various privacy concerns that emerge from the design of Android advertising libraries, mobile advertising does possess some privacy benefits relative to the web~\cite{shekhar2012adsplit}.  Because each application uses a separate instance of the advertising library, it is not possible for an advertisement or tracker to set a cookie that tracks a user across multiple applications.  (Of course, there are various unique device identifiers that can provide similar tracking.)

Upon disassembling the Google Play Services client library (version 6.5 ---December, 2014), we were able to verify that Google does not enable cookies in the WebView used by AdMob, making it somewhat more difficult for advertisers to track devices.  AdMob does, however, make use of the Android Advertising ID---a persistent device ID which can be manually reset by users, but only after a thorough search of their settings.  Additionally, the very separation of web data (tracked by cookies), and mobile data (tracked by a device ID) provide some privacy benefits---although merging multiple sources of data on a single user is not impossible.

Furthermore, apps typically provide less visibility into their internal content than web pages do.  While an app can provide keywords and other metadata through ad library API calls, most do not~\cite{book2013collusion}.  This means that an ad embedded in, for example, a music player, can target the user based on the fact that they are listening to music, but it can not target a user based on the type of music they are listening to (or build a database of the user's listening preferences) unless the app is explicitly designed to share that information with the ad library.

Examining AdMob 6.2.1 and Google Play Services 6.5, we observed that JavaScript was enabled, which is not surprising, given that many of the ads that we observed contained JavaScript content.  However, we did not observe any hooks enabled by addJavascriptInterface() to call native code.  This protects AdMob from a bug in Android that allows JavaScript to execute arbitrary Java code through reflection on versions of Android before 4.2~\cite{mwrWebViewJavaScript}.

On the flip side, the nature of mobile devices presents some new privacy concerns.  Because a user typically carries their mobile device with them throughout the day, a mobile device introduces new levels of user tracking.  Even without a GPS, this process can potentially leak much more information about a user.  With GPS, the degree of detail is much higher.  Likewise, while a user might move between various web browsers in the course of a day---for example, when they move from home to work---the mobile device remains always with them.  

Additionally, the nature of mobile tracking IDs make it easier for marketers to correlate data from different sources.  Most web tracking is done through cookies or similar unique identifiers placed on the user's computer when they visit a web page~\cite{metwalleyonline}.  Thus, there is no easy way to correlate data from two different datasets, keyed to two different sets of cookies.  However, when a tracking ID is embedded in the device, it becomes much simpler to correlate two datasets and build more complete user profiles.

In short, while there are nuances in detail between the possibilities of user targeting on the web and on mobile devices, both have significant user privacy implications.  As more and more of peoples' time and lives move into the digital world, these privacy issues will be of growing concern.

\section{Conclusions}

This work is the first of which we are aware that attempts to quantify ad targeting in mobile advertising.  We have been able to show that a large portion of mobile ads are targeted based on application, user location, time of day, and profiles built around actual users.  While our observations are limited to ads displayed through the AdMob library, this represents the largest single source of Android advertisements.  We believe that other advertising libraries and operating systems would show similar behaviors.

While our research was not intended to discern whether ads are targeted based on specific user identities, we did find that detailed user profiles appeared to be involved in the targeting of mobile advertising.  Given that mobile ads are associated with a unique device ID, it becomes clear that there are significant privacy implications to the collection of data related to the targeting of mobile ads.

Given the steady progress of technology, it is reasonable to expect user profiles to become ever more accurate and personal, making the distinction between personal identification and a theoretically anonymous profile more and more difficult to sustain.  Given the known ability to infer detailed personal information given a large enough data set of relatively nonsensitive information, we believe that the challenge of targeting can not be addressed by limiting the type of data collected, but is an intrinsic characteristic of the collection of large user datasets~\cite{chittaranjan2013mining}.

An additional issue is the question of who should have access to profiles regarding a given individual.  Internet advertising has the potential to leak profile data to various third parties~\cite{olejnik2014selling, castelluccia2012betrayed, lecuyer2014xray}.  However, there are also technical solutions that can limit data sharing while maintaining targeted advertising~\cite{toubiana2010adnostic}.

In this context, there is a need for a greater awareness of the privacy implications of mobile advertising, which we hope would lead to a greater social consensus regarding which personal information should be private from whom.  On the basis of such a consensus, society will need to take appropriate steps to manage the collection and use of personal information, both in advertising and beyond.

\section{Future Work}

There are a number of ways that we would like to extend our current research in the future.  At the most basic level, we would be able to provide greater precision in our results by engaging in larger scale data collection.  The ideal data set would include all mobile ads shown for a given period of time.  Such a dataset could be approximated by collecting data from an Internet backbone, as ads are sent in clear text, but there are obvious ethical and logistical issues to doing so.

While it is probably not possible to build a complete data set, we would like to build a better one.  That would require more collectors running for a shorter period of time, so that we could better minimize unwanted temporal effects.  We would like to have a more carefully arranged geographical distribution of our collectors, with several clusters in neighborhoods and cities with different demographic characteristics.  Likewise, with a sufficient number of collectors, we could vary device types and cellular networks which may also influence ad targeting.  That would enable us better to measure geographic effects of targeting, and to infer targeting based on income, ethnicity, and other factors.

Additionally, we would like to measure the accuracy of ad targeting by doing it ourselves.  With a well structured collection network as described above, we could target our own ads at specific demographics, and observe the extent to which our targeted ads are received by our collectors.  This would give us a greater degree of ground truth as to how effective targeting actually is.

Finally, we would like to obtain access to the data stream from an advertising exchange.  Doing so would enable us to quantify our intuition that detailed user profiles can be built from ad exchange data without ever needing to place an ad.  We would then be able to compare actual device advertising IDs with the profiles that we had inferred, in order to measure our success at compromising user privacy.

\section{Acknowledgments}

{\footnotesize \bibliographystyle{acm}
\bibliography{bibliography}}

\end{document}